\documentclass[prb,letterpaper,twocolumn,showpacs]{revtex4} 
\usepackage{times,xspace} 
\usepackage{amsbsy,amssymb,amsmath,bm,bbold} 
\usepackage{graphicx,color,epsfig,rotate} 
\usepackage{fancyhdr} 

\usepackage[normalem]{ulem}

\def\bbbc{{\mathchoice {\setbox0=\hbox{$\displaystyle\rm C$}\hbox{\hbox 
to0pt{\kern0.4\wd0\vrule height0.9\ht0\hss}\box0}} 
{\setbox0=\hbox{$\textstyle\rm C$}\hbox{\hbox 
to0pt{\kern0.4\wd0\vrule height0.9\ht0\hss}\box0}} 
{\setbox0=\hbox{$\scriptstyle\rm C$}\hbox{\hbox 
to0pt{\kern0.4\wd0\vrule height0.9\ht0\hss}\box0}} 
{\setbox0=\hbox{$\scriptscriptstyle\rm C$}\hbox{\hbox 
to0pt{\kern0.4\wd0\vrule height0.9\ht0\hss}\box0}}}}

\definecolor{purple}{cmyk}{ 0.5, 0.7, 0,0}

\begin{document} 
\title{Phase Diagram and Magnetic Excitations of Anisotropic Spin-One Magnets} 
\author{Zhifeng Zhang$^1$,  Keola Wierschem$^1$, Ian Yap$^1$,Yasuyuki Kato$^2$, Cristian D. Batista$^2$ and Pinaki Sengupta$^1$} 
\affiliation{$^1$ School of Physical and Mathematical Sciences, Nanyang Technological University, Singapore}
\affiliation{$^2$ T-Division and CNLS, Los Alamos National Laboratory, Los Alamos, NM 87545}
 
\date{\today}

\begin{abstract} 
We use a generalized spin wave approach and large scale quantum Monte Carlo  (QMC)
simulations to study the quantum phase diagram and quasiparticle excitations  of the $S=1$ Heisenberg model with an easy-plane
single-ion anisotropy  in dimensions $d=2$ and 3.  We consider two alternative approximations for describing the quantum paramagnetic state:
the standard Holstein-Primakoff approximation and a modified treatment in which the local constraint (finite dimension of the local Hilbert space)
is enforced by introducing a Lagrange multiplier. While both approximations produce qualitatively similar results, the  latter approach is the only one 
that is in good quantitative agreement with the  quantum phase diagram and the quasiparticle dispersions obtained with QMC.
This result is very important for low-temperature studies of quantum paramagnets in magnetic fields because it shows that a simple modification of the standard analytical approach should  produce  much better quantitative agreement between theory and experiment. 
\end{abstract} 
 
\pacs{75.10.Jm, 75.40.Mg, 75.40.Cx} 
 
\maketitle %
\thispagestyle{fancy}

\section{Introduction}

Lately there has been a renewed interest in the study of magnetic field induced quantum phase transitions in 
spin-one magnets with strong single-ion and exchange anisotropies \cite{Zapf06,Zvyagin07,Zapf08,Chiatti08,
Yin08,Kohama11,Zapf11,Weickert12}. The discovery of $S=1$  compounds, such as Y$_2$BaNiO$_5$ or the 
organo-metallic frameworks [Ni(C$_2$H$_8$N$_2$)$_2$(NO$_2$)]ClO$_4$ (NENP),  
[Ni(C$_2$H$_8$N$_2$)$_2$Ni(CN)$_4$] (NENC) and  [NiCl$_2$-4SC(NH$_2$)$_2$] (DTN),
fuelled experimental and theoretical studies of the role of dimensionality and single-ion anisotropy \cite{Zapf06,
Chiatti08,Yin08,Kohama11,Zapf11,Weickert12,Glarum91,Ramirez94,Sakaguchi96,Xu96,Batista98}. 
In most of the known $S=1$ magnets, the ubiquitous Heisenberg exchange is complemented 
by single-ion anisotropy. The interplay between these interactions with  external magnetic field 
and lattice geometry can result in a rich variety of quantum phases 
and phenomena, including the Haldane phase of quasi-1D systems \cite{Haldane83},  field induced Bose Einstein 
condensation (BEC) of magnetic states \cite{Zapf06,Zvyagin07,Zapf08,Chiatti08,Yin08,Kohama11,Zapf11,Weickert12} 
and field induced ferronematic ordering \cite{Wierschem12}.  Interest in $S=1$ Heisenberg antiferromagnets with 
uniaxial exchange and single-ion anisotropies has gained additional impetus recently 
after it was shown to  exhibit the spin analog of the elusive 
supersolid phase on a lattice over a finite range of magnetic fields.~\cite{Sengupta07,Sengupta07b,Peters09}

In contrast to its classical counterpart ($S \to \infty$), $S=1$ systems become {\it quantum paramagnets} (QPM) for sufficiently strong easy-plane  single-ion anisotropy. In other words, 
they  do not order down to zero temperature, $T=0$, because the dominant anisotropy term, $D  \sum_{\bm r} (S^z_{\bm r})^2$ ($D>0$), forces each spin to be predominantly
in the non-magnetic  $|S^z_{\bm r}  =0 \rangle$ state: $\langle S^z_{\bm r}  =0 | S^{\nu}_{\bm r}  | S^z_{\bm r}  =0 \rangle=0$ for $\nu=\{ x, y, z \}$. The application of a magnetic field , $H$,
along the $z$-axis reduces the spin gap linearly in $H$ since the field couples to a conserved quantity (total magnetization along the $z$-axis). The gap is closed at a quantum critical point (QCP)  where the bottom of the  $S^z=1$ branch of magnetic excitations touches zero.  This QCP belongs to the BEC universality class and the gapless mode of low-energy $S^z=1$ excitations remains quadratic for small momenta, $\omega \propto k^2$, because the Zeeman term commutes with the rest of the Hamiltonian. Since the dynamical exponent is $z=2$, the effective dimension is $d+2$ and  the upper critical dimension is $d_c=2$. 
This, and analogous field-driven transitions, have been widely studied experimentally to demonstrate BEC related phenomena in many quantum magnets. \cite{Sebastian05,Sebastian06,Zapf06,Tokiwa06,Kitada07,Yamada08,Yin08}
One of these magnets is the metal-organic framework DTN that we mentioned above \cite{Zapf06,Zvyagin07,Zapf08,Chiatti08,Yin08,Kohama11,Zapf11,Weickert12}.

The starting point of any theoretical study of a magnetic field induced phase transition in a QPM is to determine the Hamiltonian parameters, i.e., the exchange constants and the 
amplitude of the different anisotropies. The simplest  way of extracting these parameters is to fit the branches of magnetic excitations that are measured with inelastic neutron scattering (INS). The reliability of
this procedure is normally limited by the accuracy of the approach that is used to compute the dispersion relation of magnetic excitations. Numerical methods like Quantum Monte Carlo (QMC) and 
Density Matrix Renormalization Group (DMRG) are very accurate, but they can only be applied under special circumstances.  While the DMRG method \cite{White92}  has evolved to the extent that dynamical properties 
such as the frequency and momentum dependence of the magnetic structure factor can be computed very accurately \cite{Kuhner99}, its application is restricted to quasi-one-dimensional magnets such as HPIP-CuBr$_4$ \cite{Bouillot11}.
On the other hand, QMC methods can only be applied to systems that have no  frustration in the exchange interaction, i.e., that are free of the infamous sign problem.  Consequently, it is necessary to find
simple analytical approaches that are accurate enough to quantitatively reproduce the quantum phase diagram and the dispersion of magnetic excitations. 

One of the purposes of this work is to test different analytical approaches against the results of accurate QMC simulations of a spin-one Heisenberg Hamiltonian with easy-plane  single-ion anisotropy.
The model is defined either on a square or on a cubic lattice to avoid frustration and make the QMC method applicable. Besides being relevant for describing real quantum magnets, such as DTN, this model 
provides one of the simplest realizations of quantum paramagnetism and is ideal for testing methods that can be naturally extended to more complex systems.

The generic $S=1$ Heisenberg model with uniaxial single--ion  anisotropy
on an isotropic hyper-cubic lattice is given by the Hamiltonian: 
\begin{equation} 
\mathcal{H}_H = J \sum_{\langle {\bm r},{\bm r}' \rangle } {\bm S}_{\bm r} \cdot {\bm S}_{{\bm r}'}  + \! \sum_{{\bm r}} (D {S^z_{{\bm r}}}^2 - h_z S^z_{{\bm r}}) 
\label{eq:H} 
\end{equation} 
where the sum in the first term runs over nearest neighbor pairs $\langle {\bm r},{\bm r}' \rangle$. $D$ is the strength of the single ion-anisotropy, $J$ is the 
exchange constant and $h_z = g \mu_B H$, where $g$ is the g-factor and $\mu_B$ is the Bohr magneton .  
Henceforth, $J$ is set to unity and all the parameters are expressed in units of $J$. 
In this work, we shall only consider models with spatially isotropic interactions,
although the formalism can be straightforwardly generalized to anisotropic lattices.

The $(D, h_z)$ quantum phase diagram of $\mathcal{H}_H $ is well known from mean field analysis \cite{Papanicolaou90,Wang05,Hamer10}, series expansion studies\cite{Hamer08} and numerical simulations \cite{Roscilde07}. The 
$D$ term splits the local spin states into $S^z=0$ and $S^z=\pm 1$ doublet. As we explained above, the ground state  is a quantum paramagnet
for large $D\gg 1$,  i.e., it has no long range magnetic order and there is a  finite energy  gap to spin excitations. 
At finite magnetic fields, the 
Zeeman term lowers the energy of the $S^z=+1$ state until the gap closes at  a  critical field $h_c$. A canted antiferromagnetic (CAFM) phase appears right above $h_c$: the spins acquire a uniform longitudinal 
component and an antiferromagnetically ordered 
transverse component that spontaneously breaks the U(1) symmetry of global spin rotations along the $z$-axis. The CAFM phase can also be described as a condensation of  bosonic particles.  The particle density, $n_{\bm r}$, is related to the local magnetization along the symmetry axis $n_{\bm r} = S^z_{\bm r}+1$. Therefore, the magnetic field acts as a chemical potential in the bosonic description. For $h_z > h_c$, the system is populated by a finite 
density of bosons that condense in the single particle state with momentum ${\bm Q}$ with $Q_{\nu}=\pi$ ($\nu = \{ x, y, z \}$). The longitudinal magnetization (density of bosons) 
increases with field and saturates at the fully polarized (FP) state ($S_{\bm r}^z = 1 \,\,\,\forall\,\,\, {\bm r} $) 
above the  saturation field $h_s$. The FP state corresponds to a  bosonic Mott insulator 
 in the language of Bose gases. 
There exists a critical value of the single-ion anisotropy, $D_c$, below which
the CAFM phase extends down to zero field. The nature of the
QPM-CAFM quantum phase transition changes  between $h_z=0$ and $h_z \neq 0$.  The transition belongs to the BEC
universality class for $h_z \neq 0$, while it belongs to the O(2) universality class for $h_z = 0$. 

In the next section we introduce a generalized spin wave theory that describes the ground state and quasiparticle excitations 
of the quantum paramagnetic and the canted AFM phases. We describe two procedures -- one based on the 
standard Holstein-Primakoff approach \cite{Holstein40}, 
and a second one in which  a Lagrange multiplier is introduced to enforce the local constraint at a mean field 
level \cite{Sachdev90}. 
The QMC method is introduced  in Sec.~\ref{qmcm}. Sec.~\ref{ztr} includes a comparison between 
the analytical and numerical (QMC) results, which shows that  the quantitative agreement  with numerical
 simulations is considerably improved for the Lagrange multiplier method over the Holstein-Primakoff approach.
We note that this is true both for the quantum phase diagram and for the dispersion of magnetic excitations even in $d=2$. 
This remarkable accuracy in describing low energy dispersion  indicates that the second approach  
is ideally suited for extracting Hamiltonian parameters from fits of INS data.  Sec.~\ref{ftr} is devoted to finite temperature results. 
Finally, in Sec.~\ref{sum} we discuss the implication of our results for the organic quantum 
magnet DTN and for any other quantum magnet that is close to the QCP which separates the 
magnetically ordered and paramagnetic ground states.

\section{Generalized spin wave approach \label{gsw}}

In this section, we give a brief outline of the generalized spin wave formalism that was originally applied to the description of the quantum paramagnetic state of DTN \cite{Zapf06}.
Since the local Hilbert space has dimension $D_l=3$,  we introduce three Schwinger bosons (SB)
with annihilation   (creation) operators $b_{m {\bm r}}^{(\dagger)}, m\in \{0,1,2\}$. 
The three different states occupied by a single boson are mapped into the 
eigenstates of $S_{\bm r}^z$ for each site ${\bm r}$: 
\begin{equation}
b_{0 {\bm r}}^{\dag}| \emptyset \rangle=|0\rangle_{\bm r},\quad b_{1{\bm r}}^{\dag}| \emptyset \rangle=|1\rangle_{\bm r},\quad
b_{2 {\bm r}}^{\dag}|\emptyset \rangle=|-1\rangle_{\bm r}
\end{equation}
The local  constraint, 
\begin{equation}
\sum_{m=0}^2 b_{m{\bm r}}^\dagger b_{m{\bm r}} =1,
\label{eq:constraint}
\end{equation}
guarantees that the dimension of the local Hilbert space is preserved under this mapping. The bilinear forms of these SBs are generators of SU(3) in the fundamental representation \cite{Auerbach94}. We use the SBs to extend the usual SU(2) 
spin wave approach to SU(3) \cite{note1} since the local order parameter for $S=1$ spins has 8 components, which correspond to the 8 generators of the 
SU(3) group of  unitary transformations in the local Hilbert space of dimension 3. Three of them correspond to the local magnetization $(S^x_{\bm r}, S^y_{\bm r},  S^z_{\bm r})$, while the other five are the components of the traceless symmetric tensor, ${\cal Q}^{\eta \nu}_{\bm r}=(S^{\eta}_{\bm r} S^{\nu}_{\bm r} + S^{\nu}_{\bm r} S^{\eta}_{\bm r})/2-\delta_{\eta \nu}2/3$, that defines the local spin nematic moment. In particular, the paramagnetic mean field ground state  has a net nematic component induced by the single-ion anisotropy, but no net magnetization component. Such a state has no classical counterpart. Nevertheless, we can still implement a semi-classical approximation if we generalize the traditional spin-wave analysis from SU(2) to SU(3). In this approach, we can describe the quantum fluctuations around the mean field state  as small (quadratic) oscillations of an SU(3) order parameter.

At the mean field level, any ground state that is stabilized for $D > 0 $ is described  by the product state
\begin{equation}
|\psi_{cl}\rangle = \prod_{\bm r} {\tilde b}_{0 {\bm r}}^\dagger| \emptyset \rangle ,
\label{mfs}
\end{equation}
where 
\begin{equation}
{\tilde b}^{\dagger}_{0{\bm r}}=b^{\dagger}_{0{\bm r}}\cos\theta + (b^{\dag}_{1{\bm r}} \sin\theta\cos\phi + b^{\dag}_{2{\bm r}}\sin\theta\sin\phi) e^{i {\bm Q} \cdot {\bm r}}
\label{varst}
\end{equation}
and the variational parameters $\theta$ and $\phi$  are determined by minimization of the mean field energy per site 
$e_0~=~\langle  \psi_{cl}| {\cal H}_{H} |\psi_{cl}\rangle/N$:
\begin{equation}
{\partial e_0\over \partial\theta}=0,\quad\quad {\partial e_0\over \partial\phi}=0.
\label{mfmin}
\end{equation}
We note that the variational parameters $\theta$ and $\phi$ are enough to parametrize the three different phases that appear in the phase diagram of ${\cal H}_{H}$ for $D>0$. 
The bosonic operator ${\tilde b}^{(\dagger)}_{0 {\bm r}}$ belongs to a new set of SB operators that are obtained from the original set $\{b_{m {\bm r}}^{(\dagger)}\}$
by a unitary transformation, ${\cal U}_{\bm r}$:
\begin{equation}
{\tilde{\bm b}}_{\bm r} = {\cal U}_{\bm r} {\bm b}_{\bm r},
\quad
{\bm b}_{\bm r} = \left (
\begin{array}{c}
b_{0 {\bm r}} \\
b_{1 {\bm r}}\\
b_{2 {\bm r}}.
\end{array}
\right ).
\end{equation}
This transformation corresponds to choosing a quantization axis along the direction of the order parameter, as it is done
in the usual spin wave treatment. Since the ground state of the antiferromagnetic phase 
breaks translational symmetry making the two sublattices inequivalent, the corresponding canonical transformation, ${\cal U}_{\bm r}$, is
different for the two sublattices, as it is clear from the phase factor $e^{i {\bm Q} \cdot {\bm r}}$ that appears in Eq.\eqref{varst}.

In terms of the SBs, the spin operators $S^\mu_{\bm r}$ 
assume bilinear forms $ S^\mu_{\bm r}={\bm b}_{\bm r}^\dagger{\cal S}^\mu{\bm b}_{\bm r}$, 
\begin{eqnarray}
S^x_{\bm r} &=& \frac{1}{\sqrt{2}} (b_{1{\bm r}}^\dagger b_{0{\bm r}} + b_{0{\bm r}}^\dagger b_{2{\bm r}} ),
\nonumber \\
S^y_{\bm r} &=& \frac{1}{\sqrt{2}i} (b_{1{\bm r}}^\dagger b_{0{\bm r}} - b_{0{\bm r}}^\dagger b_{2{\bm r}} ),
\nonumber \\
S^z_{\bm r} &=& b_{1{\bm r}}^\dagger b_{1{\bm r}} -  b_{2{\bm r}}^\dagger b_{2{\bm r}},
\end{eqnarray}
that transform  as ${\tilde S}^\mu_{\bm r}={\cal U}_{\bm r} S^\mu_{\bm r}{\cal U}^\dagger_{\bm r}$. The spatial dependence of the unitary transformation ${\cal U}_{\bm r}$
can be eliminated if we change the original basis of the Hamiltonian ${\cal H}_H$. In particular, the CAFM state becomes uniform if we rotate the spin reference frame of one of the sublattices by angle $\pi$ along the $z$-axis. Since the uniform paramagnetic ground states of ${\cal H}_H$ remain invariant under this transformation, the unitary transformations ${\cal U}_{\bm r}$ become ${\bm r}$-independent in the new basis for all the different phases of ${\cal H}_H$. Since, $S^z_{\bm r} \to S^z_{\bm r}$ and $S^{x,y}_{\bm r} \to -S^{x,y}_{\bm r}$, we have that ${\cal H}_H \to$
\begin{equation} 
{\cal H}_H = J \sum_{\langle {\bm r},{\bm r}'  \rangle, \nu}  a_{\nu} S^{\nu}_{\bm r}  S^{\nu}_{{\bm r}'}  + \! \sum_{{\bm r}} (D {S^z_{{\bm r}}}^2 - h_z S^z_{{\bm r}}) 
\label{eq:H2} 
\end{equation} 
in the new basis, where $a_z=1$ and $a_x=a_y=-1$. We note that this change of basis shifts the AFM wave vector from ${\bm Q}$ to ${\bm 0}$ and removes the 
factor $e^{i {\bm Q} \cdot {\bm r}}$ from Eq.~\eqref{varst}.

The bosonic representation of the Hamiltonian in the new basis is
\begin{eqnarray}
{\cal H}_H &=&  J \sum_{\langle {\bm r}, {\bm r}' \rangle,\nu} a_{\nu}  {\tilde{\bm b}}^\dagger_{\bm r} {\tilde{\cal S}}^\nu{\tilde{\bm b}}_{\bm r} {\tilde{\bm b}}^\dagger_{\bm r'}{\tilde{\cal S}}^\nu{\tilde{\bm b}}_{\bm r'}  \\ \nonumber
&+& D\sum_{\bm r} \left ( 1- {\tilde{\bm b}}^\dagger_{\bm r} {\tilde{\cal A}}{\tilde{\bm b}}_{\bm r} \right ) - h_z\sum_{\bm r} {\tilde{\bm b}}^\dagger_{\bm r} {\tilde{\cal S}}^z{\tilde{\bm b}}_{\bm r}
\label{eq:Hb}
\end{eqnarray}
where 
\[
{\tilde {\cal S}}^\mu = {\cal U}{\cal S}^\mu{\cal U}^\dagger, \quad{\tilde {\cal A}} = {\cal U}{\cal A}{\cal U}^\dagger\quad\mbox{and}\quad {\cal A}_{ij}=\delta_{i0}\delta_{j0}
\]
 
The condensation of the bosons ${\tilde b}_{0 {\bm r}}$  is implemented via a natural extension of the  Holstein-Primakoff transformation~\cite{Holstein40} to the case of 
more than one type of boson. From the local  constraint (\ref{eq:constraint})  we obtain:
\begin{equation}
\tilde{b}_{0 {\bm r}}^\dagger =  \tilde{b}_{0 {\bm r}} =
\sqrt{1-\tilde{b}_{1 {\bm r}}^\dagger \tilde{b}_{1 {\bm r}}-\tilde{b}_{2 {\bm r}}^\dagger \tilde{b}_{2 {\bm r}}}
\label{eq:bec}
\end{equation}
By applying the above condition to the Hamiltonian (\ref{eq:H}) and keeping terms up to bilinear in the
bosonic creation and annihilation operators, we obtain the mean field ground state energy
\begin{equation}
e_0=d J\sum_\nu   a_{\nu} \tilde{S}_{00}^\nu\tilde{S}_{00}^\nu -h_z\tilde{S}_{00}^z+D(1-\tilde{A}_{00})
\end{equation}
and the spin wave Hamiltonian
\begin{eqnarray}
{\cal H}_{sw}  &=& \hspace*{-0.4cm} \sum_{\begin{subarray}{c}\langle {\bm r}, {\bm r}' \rangle\\\alpha,\beta\in\{1,2\}\end{subarray}}\hspace*{-0.4cm} 
\left[  t_{\alpha\beta} {\tilde b}^\dagger_{\alpha{\bm r} }{\tilde b}_{\beta {\bm r}' } 
+ \Delta_{\alpha\beta} {\tilde b}^\dagger_{\alpha {\bm r} }{\tilde b}^\dagger_{\beta {\bm r}' } + {\rm H. c.} \right] 
\nonumber \\
&+& \hspace*{-0.4cm} \sum_{\begin{subarray}{c}{\bm r}\\\alpha,\beta\in\{1,2\}\end{subarray}
}\hspace*{-0.4cm} \lambda_{\alpha\beta}{\tilde b}^\dagger_{\alpha {\bm r} }{\tilde b}_{\beta {\bm r}}
\label{eq:hsw}
\end{eqnarray}
with the Hamiltonian parameters  
\begin{eqnarray}
\nonumber t_{\alpha\beta} &=&   J\sum_\nu  a_{\nu} {\tilde{\cal S}}^\nu_{\alpha 0}{\tilde{\cal S}}^\nu_{0\beta} \\
\nonumber \Delta_{\alpha\beta} &=&  J\sum_\nu  a_{\nu} ({\tilde{\cal S}}^\nu_{\alpha 0}{\tilde{\cal S}}^\nu_{\beta 0} - ({\tilde{\cal S}}^\nu_{00})^2\delta_{\alpha\beta})\\
\lambda_{\alpha\beta} &=& d J\sum_\nu  a_{\nu} {\tilde{\cal S}}^\nu_{\alpha \beta}{\tilde{\cal S}}^\nu_{00} + D\delta_{\alpha\beta} -h_z{\tilde{\cal S}}^z_{\alpha\beta}
\end{eqnarray}
where $d$ is the spatial dimension.
In the next step, the spinwave Hamiltonian (\ref{eq:hsw}) is transformed to momentum 
representation by introducing bosonic operators in momentum space:
\begin{equation}
{\cal H}_{sw} =\sum_{{\bm k},\alpha,\beta} \epsilon_{\alpha\beta}({\bm k}) {\hat b}_{\alpha{\bm k}}^\dagger {\hat b}_{\beta{\bm k}}
+ { \gamma_{\alpha\beta}({\bm k})\over 2}\left ({\hat b}_{\alpha{\bm k}}^\dagger {\hat b}_{\beta -{\bm k}}^\dagger + {\rm  H. c.  }  \right ),
\end{equation}
with
\begin{eqnarray}
{\hat b}_{\alpha{\bm k}}^\dagger &=& \frac{1}{\sqrt{N}} \sum_{\bm r} e^{{\bm k} \cdot {\bm r}} {\tilde b}^\dagger_{\alpha {\bm r} },
\nonumber \\
\epsilon_{\alpha\beta}({\bm k}) &=& \lambda_{\alpha \beta} + t_{\alpha\beta} \sum_{\nu}  \cos{k_{\nu}}
\nonumber \\
\gamma_{\alpha\beta}({\bm k}) &=&  \Delta_{\alpha\beta} \sum_{\nu}  \cos{k_{\nu}}
\end{eqnarray}
The resultant Hamiltonian can then be straightforwardly diagonalized by a Bogoliubov transformation 
to yield the single particle dispersion:
\begin{equation}
{\cal H}_{sw} =\sum_{{\bm k},\alpha} \omega_{{\bm k} \alpha} 
\left ( a_{\alpha {\bm k}}^\dagger a^{\;}_{\alpha{\bm k}} + \frac{1}{2} \right ) - \frac{\epsilon_{\alpha\alpha}({\bm k})}{2}
\label{gswH}
\end{equation}

\subsection{QPM phase and the Fully Polarized phase}

At the mean field level, the paramagnetic state, 
\begin{equation}
|\psi_{cl} (\theta=0) \rangle = \prod_{\bm r} b_{0{\bm r}}^\dagger|\emptyset\rangle ,
\end{equation}
is the lowest energy state  for large enough $D$, as long as the applied magnetic field remains below a critical value $h_c$.
Since  the unitary transformation can be chosen as the identity, ${\cal U} =  \mathbb{1}$,  the quasiparticle dispersion becomes particularly 
simple in  the QPM phase:
\begin{equation}
\omega_{{\bm k}\pm}=\sqrt{D^2+2D\eta_{\bm k}} \pm h_z,\quad \eta_{\bm k}=-2J\sum_\nu\cos(k_\nu).
\label{qpmdisp}
\end{equation}
Both branches have the same dispersion at zero field, $h_z=0$, as expected from  time reversal symmetry.
A finite $h_z$ splits the branches linearly in $h_z$ without changing the dispersion. This is a consequence of the fact that the external field couples to
the total magnetization, $M^z= \sum_{\bm r} S^z_{\bm r}$, which is a conserved quantity. Both branches have a minimum at  the AFM wave-vector ${\bm k}={\bm 0}$ that determines the size of the gap. The dispersion is quadratic near  ${\bm k}={\bm 0}$ except for the critical point $(D_c=4dJ, h_z=0)$ that separates the QPM phase from the CAFM phase at $h_z=0$. 
The field induced QCP then belongs to the BEC universality class in dimension $d+2$.  By expanding around ${\bm k}={\bm 0}$, we obtain:
\begin{equation}
\omega_{{\bm k}\pm}\approx J k^2\sqrt{D/(D-D_c)}+\sqrt{D(D-D_c)}\pm h_z
\label{eq:wkQPM}
\end{equation}
It is clear from this expression that the effective mass of the magnetic excitations vanishes for $D \to D_c$: $m^* \propto \sqrt{D-D_c}$. This is indeed the expected behavior if
we keep in mind that the dispersion must be linear at the the critical point $(D_c=4dJ, h_z=0)$ ($z=1$ for the O(2) QCP as we discussed in the introduction).

The QPM ground state remains stable for
\begin{eqnarray}
D & \ge &  D_c=4 d J \nonumber \\
h_z & \le & h_c = \sqrt{D(D-D_c)}
\label{sc}
\end{eqnarray}
The ground state becomes fully polarized over the saturation field
\begin{equation}
h_s=D+4 d J,
\label{hsat}
\end{equation}
and the mean field state,
\begin{equation}
| \psi_{cl}(\theta=\pi/2,\phi=0) \rangle=\displaystyle\prod_{\bm r} b_{1{\bm r}}^{\dag}|\emptyset \rangle, 
\end{equation}
coincides with the exact ground sate. 
The energy of the system is proportional to the applied field as expected. The two branches of magnetic excitations above the saturated state are given by:
\begin{eqnarray}
\omega_{{\bm k}1} &= & h_z - D -2 d J + \eta_{\bm k},
\label{dispsat1}
\nonumber \\
\omega_{{\bm k} 2} &= & 2 h_z.
\label{dispsat2}
\end{eqnarray}
The flat branch, $\omega_{k2}$, describes the approximated spectrum of two-magnon bound states that appear above a critical value of the single-ion anisotropy \cite{Zvyagin07}.

By comparing Eqs.\eqref{qpmdisp} and \eqref{dispsat2}, we can see that the masses of the gapless bosons at the two field induced QCPs, $h=h_c$ and $h=h_s$ can be very 
different: 
\begin{eqnarray}
\frac{1}{m^*}  &=& \frac {\partial^2 \omega_{{\bm k} -}}{\partial k ^2} |_{{\bm k}={\bm 0}} =2 J \sqrt{D/(D-D_c)} ,
\nonumber \\
\frac{1}{m}  &=& \frac {\partial^2 \omega_{{\bm k} -}}{\partial k ^2} |_{{\bm k}={\bm 0}} = 2 J.
\end{eqnarray}
While the  mass renormalization factor $m^*/m = \sqrt{(D-D_c)/D}$ may not be quantitatively accurate, the obtained mean field critical exponent, 
$m^*/m \propto \sqrt{(D-D_c)/D}$, is correct for $d=3$ up to logarithmic corrections, because $d_c=3$ is the upper critical dimension for the O(2) QCP in 
dimension $d+1$. For $d\leq d_c$ we have 
\begin{equation}
m^*/m \propto \Delta_s \propto (D-D_c)^{\nu z}
\end{equation}
 and the mean field exponent $\nu=1/2$ is not correct for $d<3$. 
It is clear then that quantum paramagnets which are close to the CAFM instability ($D \gtrsim D_c$) should exhibit a very large asymmetry between the mass
of the bosonic excitations for $h \leq h_c$ and $ h \geq h_s$. This is indeed the case of the compound DTN whose thermodynamic properties exhibit a 
large asymmetry between the the two critical points at $h_c$ and $h_s$. The possibility of having a relatively large $m^*/m$ ratio that can be tuned with pressure
allows for measuring dependence of different physical properties on the mass of the bosonic excitations. This property of certain quantum paramagnets is particularly useful
for unveiling the dominant scattering mechanism for thermal conductivity, $\kappa$, because different mechanisms usually lead to different dependences of $\kappa$ on the mass of the quasiparticles \cite{Kohama11}.   

While the linear approach that we have described gives the correct qualitative picture in $d=3$, it is still far from being quantitatively accurate in $d=3$ or $d=2$, as we will see in the next sections. This shortcoming can be a serious problem for comparisons against experimental data. In particular, the Hamiltonian parameters for quantum paramagnets are normally extracted from fits of the quasiparticle dispersions that are measured with INS \cite{Zapf06}. The accuracy of the obtained Hamiltonian parameters depends on the accuracy of the approach that is used for computing the dispersions $\omega_{{\bm k}\nu}$. Moreover, for quantum paramagnets like DTN which have low critical fields, $h_c \ll h_s -h_c$, the linear approach normally predicts  AFM ordering at $h_z=0$. Therefore, it is necessary to modify the linear approach in order to obtain a quantitatively accurate description of the low field paramagnetic ground state and the low-energy excitations. As we shall see in the next sections, the modified approach that was originally applied to the description of DTN \cite{Zapf06} and that we describe in the rest of this subsection, is quantitatively accurate for $d=3$ and $d=2$.  

In the modified approach we replace Eq.\eqref{eq:bec} by
\begin{equation}
\langle \tilde{b}_{0 {\bm r}}^\dagger\rangle = \langle \tilde{b}_{0 {\bm r}}\rangle= s,
\label{eq:bec2}
\end{equation}
and impose the constraint  \eqref{eq:constraint}  at a mean field level by introducing the Lagrange multiplier $\mu$:
\begin{equation}
{\cal H}_H \to {\bar {\cal H}}_H = {\cal H}_H - \mu \sum_{\bm r} \left(  1- \sum_{m=0}^2 {\tilde b}_{m{\bm r}}^\dagger {\tilde b}_{m{\bm r}}   \right )
\end{equation}
The rest of the procedure is similar to spin-wave theory, 
i.e., we only keep terms up to quadratic order in the bosonic operators
${\tilde b}^{(\dagger)}_{m{\bm r}}$ ($m=1,2$) and diagonalize the resulting quadratic Hamiltonian via a Bogolyubov transformation.
This procedure leads to the diagonal form \eqref{gswH}, but with a modified quasiparticle dispersion,
\begin{equation}
\omega_{{\bm k}\pm}=\sqrt{\mu^2+2\mu s^2 \eta_{\bm k}} \pm h_z,
\label{qpmdisp2}
\end{equation} 
relative to the expression \eqref{qpmdisp} that was obtained from the linear approximation. We note that the new dispersion \eqref{qpmdisp2} can be obtained from
the previous one if we replace $D$ by $\mu$ and $J$ by $J s^2$. Therefore, in the quantum paramagnetic state, the net effect of including a Lagrange multiplier to enforce the 
constraint \eqref{eq:constraint} at the mean field level   is a renormalization of the single-ion anisotropy and exchange parameters.

The parameters $s$ and $\mu$ are determined self-consistently by the saddle point equations \cite{Sachdev90}: 
\begin{equation}
\left \langle   \frac{\partial {\bar {\cal H}}_H} {\partial \mu } \right \rangle =0, \;\;\;\;
\left \langle   \frac{\partial {\bar {\cal H}}_H} {\partial s } \right \rangle =0. 
\label{speq}
\end{equation}
By explicitly computing the left hand side of these two equations we obtain the following expressions:
\begin{eqnarray}
D = \mu \left( 1 + \frac{1}{N} \sum_{\bm k}  \frac{\eta_{\bm k}}{\sqrt{\mu^2 + 2 s^2 \mu \eta_{\bm k}}} \right),
\nonumber \\
s^2 = 2 - \frac{1}{N} \sum_{\bm k}  \frac{(\mu + s^2 \eta_{\bm k})}{\sqrt{\mu^2 + 2 s^2 \mu \eta_{\bm k}}}.
\end{eqnarray}
The stability conditions \eqref{sc} for the QPM ground state are replaced by
\begin{eqnarray}
\mu & \ge &  \mu_c=4 d s^2  J, \\
h_z & \le & h_c = \sqrt{\mu (\mu-\mu_c)}. 
\label{sc2}
\end{eqnarray}
As we will see in the next sections, the quantum phase diagram that is obtained from these modified conditions is in much better agreement with QMC simulations. 
The same is true for the modified quasiparticle dispersion \eqref{qpmdisp2}.

\subsection{Canted Antiferromagnetic (CAFM) phase}
To describe the CAFM phase, one needs to use the general expression
for the condensed boson with ${\cal U} \neq  \mathbb{1}$. In particular, we use the expression given by Eq.~\eqref{varst}
\begin{equation}
{\tilde b}^{\dag}_{0{\bm r}}=b^{\dag}_{0{\bm r}}\cos\theta + b^{\dag}_{1{\bm r}} \sin\theta\cos\phi + b^{\dag}_{2{\bm r}}\sin\theta\sin\phi .
\end{equation}
We recall that the factor $e^{i {\bm Q} \cdot {\bm r}}$ is removed from  Eq.~\eqref{varst} after the change of basis that led to Eq.~\eqref{eq:H2}.
The other bosonic operators are obtained by orthogonalization.
The parameters $\theta$ and $\phi$ are determined by the minimization of the mean field
energy [see Eq.\eqref{mfmin}].
In the absence of any applied field, the AFM ordered phase is invariant under the product of a translation by one lattice parameter and a time reversal transformation.
This symmetry implies that $\phi={\pi\over 4}$, i.e.,
the local moments have equal weights in the $S_z=\pm 1$ states. By minimizing the mean field energy as a function of the remaining variational parameter, $\theta$, 
 we obtain 
\begin{equation}
\sin^2\theta = {1\over 2} - {D\over 16 dJ}.
\end{equation}
The dispersion relation consists of two non-degenerate branches that, in the low energy
limit ($k \rightarrow 0$), are given by
\begin{eqnarray}
\omega_{{\bm k}1} &\approx & \sqrt{D_c^2-D^2} + {D^2 \over 4 d \sqrt{D_c^2-D^2}} k^2,
\nonumber \\
\omega_{{\bm k} 2} &\approx & \sqrt{J(D_c+D)} k.
\label{dcafm}
\end{eqnarray}

Unfortunately, the modified approach based on the inclusion of a Lagrange multiplier that we introduced in the previous subsection 
does not work well inside the ordered phase. Both branches become gapped inside the ordered phase, i.e., 
the  approach misses the Goldstone mode associated with the spontaneous breaking of the U(1) symmetry of global spin rotations along the $z$-axis.

As we explained above, the magnetic field induced quantum phase transition from the QPM to the 
CAFM phase is qualitatively different from the transition between the same two phases that is 
induced by a change of $D$ at $h_z=0$.  Eq.\eqref{qpmdisp} shows that the effect of increasing 
$h_z$ from zero at a fixed $D > D_c$ is to reduce the gap, 
$\Delta_s=\omega_{{\bm k}={\bm 0} -} = \sqrt{D^2 - 4 d J D}- h_z$, linearly in $h_z$. The dispersion  
does not change because $h_z$ couples to $m_z = \sum_{\bm r} S^z_{\bm r}/N$ that is a conserved 
quantity ($m_z=1$ for the spin excitations that have dispersion $\omega_{{\bm k} -}$). Therefore, the  
quasiparticle dispersion remains quadratic at the field induced QCP $h=h_c=\sqrt{D^2 - 4 d J D}$, 
i.e., the dynamical exponent is $z$=2. The field induced QCP then belongs to the BEC universality 
class in dimension $d+2$. On the other hand, if the single-ion anisotropy is continuously decreased 
at zero applied field,  the two branches remain degenerate  
and the gap vanishes at  $D = D_c$ ($h_z=0$). The low-energy dispersion becomes {\em linear} at the 
QPM-CAFM phase boundary, $\omega_{{\bm k}\pm}\approx\sqrt{ 2D J} k$ for small ${\bm k}$.  As it is 
clear from Eq.~\eqref{dcafm}, the degeneracy between the two branches at $h_z=0$  is lifted  inside the CAFM phase
 -- one of the branches, $\omega_{{\bm k}2}$, remains gapless with a linear dispersion at low 
energy (corresponding to the Goldstone mode of the ordered CAFM state ) 
whereas  the other mode develops a gap to the lowest excitation.

In the following sections, we shall use large scale quantum Monte Carlo simulations of 
the Hamiltonian (\ref{eq:H}) to demonstrate that the introduction of a Lagrange multiplier significantly improves the quantitative description of the QPM phase, and that the 
linear approximation gives a qualitatively correct description of the quantum phase transitions in $d=3$. As expected, in $d=2$, the only deviation from mean field behavior  occurs at the O(2) QCP, $D=D_c$ and $h_z=0$, because the effective dimension, ${\cal D}= d +1$, is lower than four.

\section{Quantum Monte Carlo method \label{qmcm}}

We have used two different QMC methods,
the standard stochastic series expansion (SSE)  with loop updates \cite{sse1,sse2,dloops} 
and a modified version developed in Ref.~\onlinecite{kato2009},
to study the ground state and finite temperature properties of the Hamiltonian (\ref{eq:H}). 
Since both methods are unbiased and exact within the statistical error,
we refer to them as QMC collectively in this paper.
On the dense parameter grids (temperature for thermal transitions and magnetic 
field or single-ion anisotropy for ground state transitions) needed to study 
the critical region in detail, the statistics of the QMC results can be 
significantly improved by the use of a parallel tempering scheme \cite{tempering1,tempering2}. 
The implementation of tempering schemes in the context of the SSE method has 
been discussed in detail previously \cite{ssetemp1,ssetemp2}. 
Ordinarily, the SSE would suffer from the negative sign problem for the AFM Heisenberg
interaction. However,  the sublattice rotation discussed in section II maps the
 XY part of the Heisenberg interaction into a ferromagnetic exchange term, thus alleviating 
the sign  problem. This transformation 
maps the AFM ordering vector to ${\bf Q}={\bm 0}$ in the new basis.

We compute the spin stiffness  $\rho_s$ --- defined as the response to a twist in the boundary conditions\cite{kohn,kopietz}. 
The transition to CAFM is efficiently investigated by studying the scaling properties of the spin stiffness  $\rho_s$.
For simulations that sample multiple winding number sectors, the stiffness can be related to the fluctuations 
of the winding number in the updates \cite{pollock,harada,sse1,cuccoli} 
and can be estimated readily with great accuracy. 
For the isotropic systems that are primarily considered in the present study, the estimates of the stiffness along all the axes are equal within statistical fluctuations.
Along with the spin stiffness, we calculate the square of the order parameters characterizing 
the different ground states as well as standard thermodynamic observables 
such as energy and magnetization.
The transverse component of the spin static structure function, 
\begin{equation} 
 S^{+-}({\bm q})={1\over N}\sum_{{{\bm r}},{\bm r}'}e^{-i{\bm q}\cdot({\bm r}-{\bm
r}')} \langle S^+_{\bm r}S^-_{{\bm r}'} \rangle,
\end{equation} 
measures the off-diagonal long-range ordering in the XY plane. Its
value at the AFM ordering wave vector, $S^{+-}_{\bm Q}$, quantifies the
XY AFM order. In the bosonic language, it is the condensate fraction of
the BEC.  We also compute the  mean value of the $zz$-component of the nematic  tensor  component,
${\cal Q}^{zz}_{\bm r}=\langle (S_{\bm r}^z)^2-{2\over 3}\rangle $, that is induced by the single-ion anisotropy term.

\section{Zero-temperature results \label{ztr}}

\subsection{Finite-size scaling for quantum criticality}
The continuous phase transition from the QPM phase to the CAFM phase is marked by the closing of the spin gap. 
To determine the transition point, we use the finite-size scaling properties of the spin stiffness $\rho_s$.
 The finite-size scaling analysis at the critical point predicts that 
\begin{eqnarray}
    \rho_s(L, \beta,D)
    &\sim& L^{2-d-z} Y_{\rho_s}(\beta/L^z,(D-{D}_c)L^{1/\nu}),
    \nonumber
\end{eqnarray}
below the upper critical dimension, i.e., $d+z\leq 4$,
where  $L$ is the linear dimension of the system,  $z$ is the dynamic critical exponent,
and $Y_{\rho_s}$ is the scaling function. 
$z=1$  for QPTs belonging to the O(2) universality class and $z=2$ for BEC QCPs.
Since the effective dimension of  the BEC-QCP in $d=3$, ${\cal D}= 3 +2$, is above the upper critical dimension ${\cal D}_c=4$,
we need to apply a modified finite-size scaling \cite{kato2010}
\begin{eqnarray}
    \rho_s(L, \beta, h_z )
    &\sim& L^{-(d+z)/2} Y_{\rho_s}(\beta/L^z,(h_z-h_c)L^{(d+z)/2}).
    \nonumber
\end{eqnarray}

The scale invariance at the critical point provides a powerful and widely used tool to simultaneously determine 
the position of the critical point and verify the value of $z$. 
On a plot of $\rho_s L^{d+z-2}$ or $\rho_s L^{(d+z)/2}$ as a function of the driving parameters, $D$ or $h_z$, 
the curves for different system sizes will cross at the critical point provided 
the correct value of $z$ is used.

\begin{figure}[htb]  
\includegraphics[width=0.9\textwidth, clip]{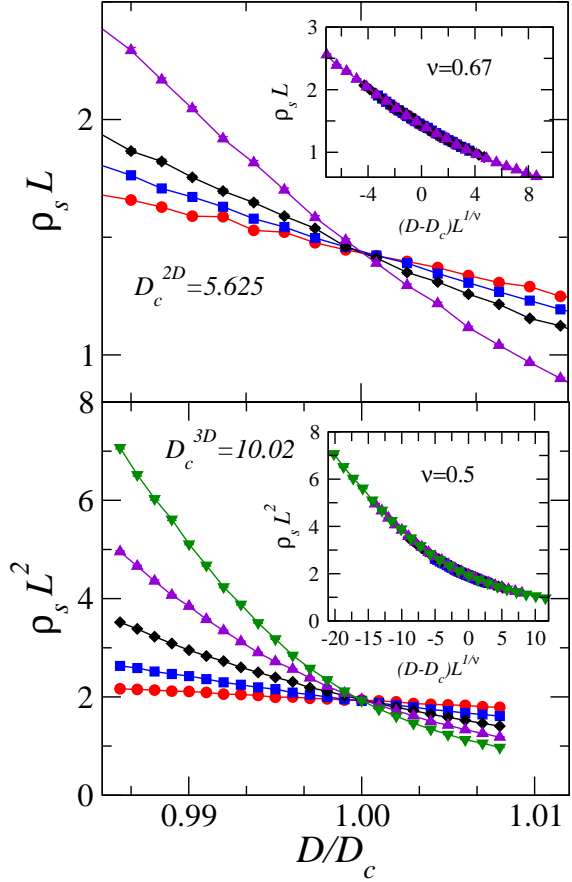} 
\caption{(Color online) 
Finite-size scaling plots of spin stiffness $\rho_S$. 
The four system sizes of the square lattices (upper panel) $L\times L$ are $8\times8$ (red), $10\times10$ (blue), $12\times12$ (black) and $18\times18$ (purple). The five system sizes of the cubic lattices (lower panel) $L\times L\times L$ are $4\times4\times4$ (red), $6\times6\times6$ (blue), $8\times8\times8$ (black), $10\times10\times10$ (purple) and $12\times12\times12$ (green). The temperatures are taken to be $T=L/4$ in the square lattice and $L/2$ in the cubic lattice.
The boundary conditions are periodic.
} 
\label{fig:Dc} 
\end{figure} 

Figs.~\ref{fig:Dc}  shows the scaling of the stiffness close to the critical point for the
QPM-CAFM transition at $h_z=0$ driven by varying the single-ion anisotropy $D$. 
From field theoretic arguments, the transition is expected to belong to the O(2) universality class for which $z=1$. 
Indeed, the curves were found to exhibit a unique crossing point only for $z=1$. 
For a square lattice (top panel), we obtain a critical $D_c = 5.63$, in agreement with previous results~\cite{Roscilde07} , whereas the
transition occurs at $D_c=10.02$ on a cubic lattice (bottom panel). 
Further confirmation of the O(2) universality class of the transition is shown in the inset panels where on
 a plot of $\rho_s L^{d+z-2}$ vs. $(D-D_c) L^{1/\nu}$,
 the data for different system sizes collapse onto a single curve with our estimated $D_c$
 and known critical exponents for the O(2) universality class in $d+1$ dimensions.

\begin{figure}[hpt] 
\includegraphics[width=8cm]{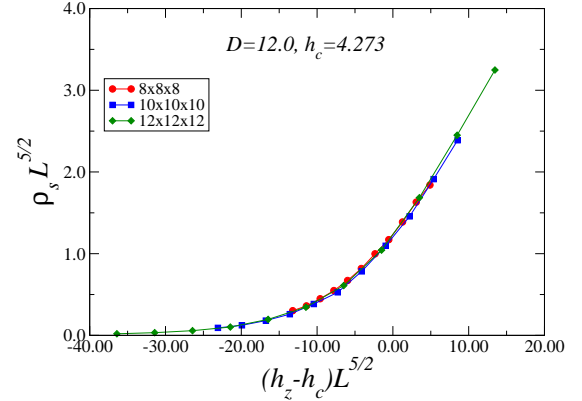} 
\caption{(Color online) Determination of the critical field through finite size scaling with $z=2$ that confirms the BEC universality class of the field induced quantum
critical points.} 
\label{fig:D12} 
\end{figure} 

Fig.~\ref{fig:D12} shows the modified finite-size scaling plots of the QPM to CAFM transition for $D > D_c$ as the field $h_z$ is varied. 
The transition is expected to belong to the BEC universality class 
and scale invariance for the stiffness at the critical point is found for $z=2$
in accordance with field theoretic predictions. 
Thus the analysis of the stiffness data at the quantum critical
points show that the QPM -- CAFM transition belongs to the O(2) universality
class for $h_z=0$, but changes to BEC universality class for $h_z \neq 0$. 

\subsection{Quasiparticle dispersion in the QPM phase \label{quadisp}}

\begin{figure}[hpt] 
\includegraphics[width=8cm, trim=30 30 50 30,clip]{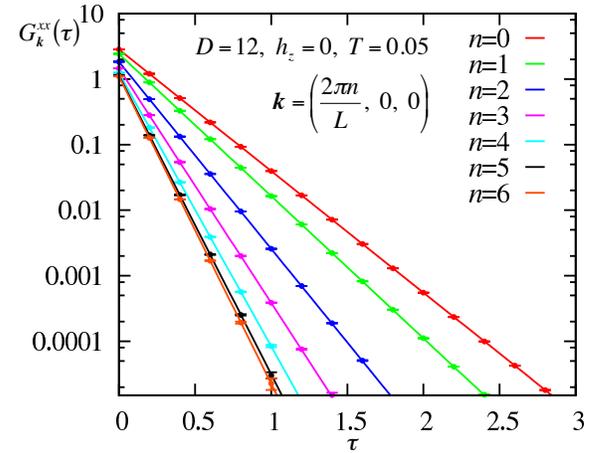} 
\caption{
  (Color online) Imaginary time Green's function computed with QMC for $D=12$, and $h_z=0$. The linear size of the 
  finite cubic lattice is $L=12$ and the boundary conditions are periodic. The solid fitting lines correspond to the function 
  defined in Eq.~\eqref{fitfun}.} 
\label{fig:Gt} 
\end{figure} 

\begin{figure}[hpt] 
\includegraphics[width=8cm, trim=40 30 390 30,clip]{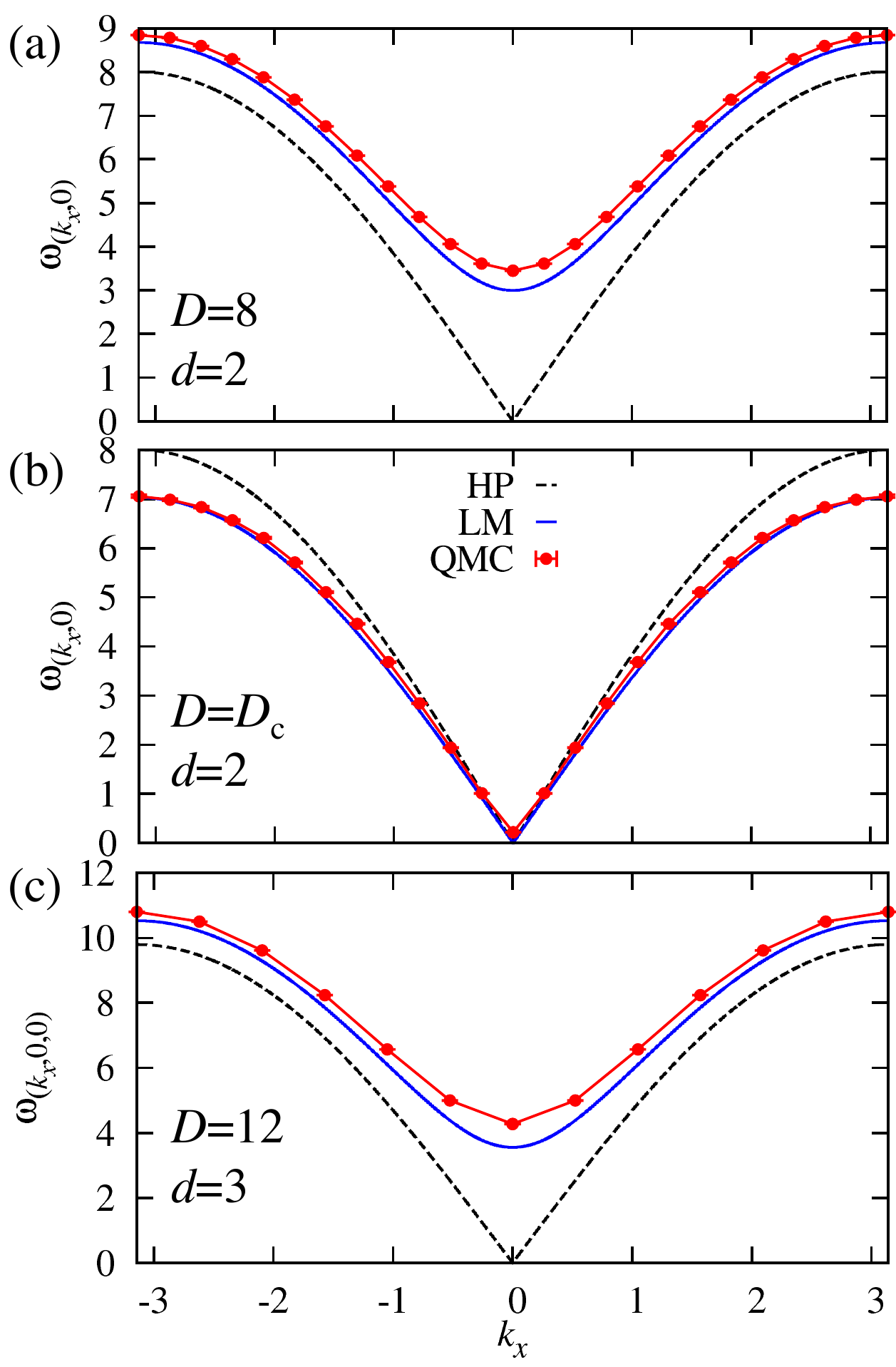} 
\caption{(Color online) Dispersions of the single magnon excitation (a) $D=8$ in 2D, (b) $D=D_c$ in 2D and (c) $D=12$ in 3D.
In 2D, $D_c=$ 8, 5.71 and 5.625 for the linear HP, LM and QMC approaches, respectively. } 
\label{fig:disp} 
\end{figure} 

The phase boundary between QPM and CAFM phases is also determined by the value of the single magnon excitation gap $\Delta_s$.
Since the Zeeman term commutes with the rest of the Hamiltonian, the spin gap of the QPM phase changes linearly in the magnetic field and vanishes at the critical field
$h_c = \Delta_{s}(h_z=0)$. The quasiparticle  dispersion and the gap $\Delta_{s}$ can be extracted from the QMC results by analysing 
the imaginary time Green's function
\begin{eqnarray}
G^{xx}_{\bm k}(\tau) &=&\frac{1}{L^d}\sum_{\bm r} 
\left\langle S^x_{\bm r}(\tau) S^x_{\bm 0}(0)\right\rangle e^{i{\bm k}\cdot{\bm r}}.
\end{eqnarray}
The quasiparticle dispersion is computed 
by fitting the QMC data of $G^{xx}_{\bm k}(\tau)$ with the function
\begin{eqnarray}
f( \tau ) =
 A \left[ e^{-\omega \tau} + e^{-\omega (\beta - \tau)} \right],
 \label{fitfun}
\end{eqnarray}
where $A$ and $\omega$ are fitting parameters. In particular, the parameter $\omega$ corresponds to the magnetic excitation energy for each momentum ${\bm k}$.
Figure \ref{fig:Gt} shows that the fit is nearly perfect for the $G^{xx}_{\bm k}(\tau)$ curve that is obtained  in the QPM phase.
The estimated phase boundary is  $h_c=4.2726(3)$ for $D=12$, $d=3$ and $L=12$.
This estimation is fully consistent with the modified finite-size scaling analysis. (See Fig.~\ref{fig:D12}.)
Since finite size effects are very small deep inside the QPM state (far from critical point),
the field induced phase boundary can be estimated very precisely with $L=12$.
Fig.~\ref{fig:disp} shows the comparison between the quasiparticle  dispersions obtained from the QMC results and the analytical expressions \eqref{qpmdisp} and \eqref{qpmdisp2} that we derived in the previous section using the Holstein-Primakoff  (HP) and the Lagrange multiplier (LM)  approaches.
The quantitative agreement with the numerical result is much better for the  LM approach that reproduces not only the value of the spin gap and the overall dispersion inside the 
QPM phase, but also the spin velocity at the O(2) QCP $D=D_c (h_z=0)$.

\subsection{Quantum phase diagram}
\begin{figure}[hpt] 
\includegraphics[width=8cm, trim=30 50 30 30,clip]{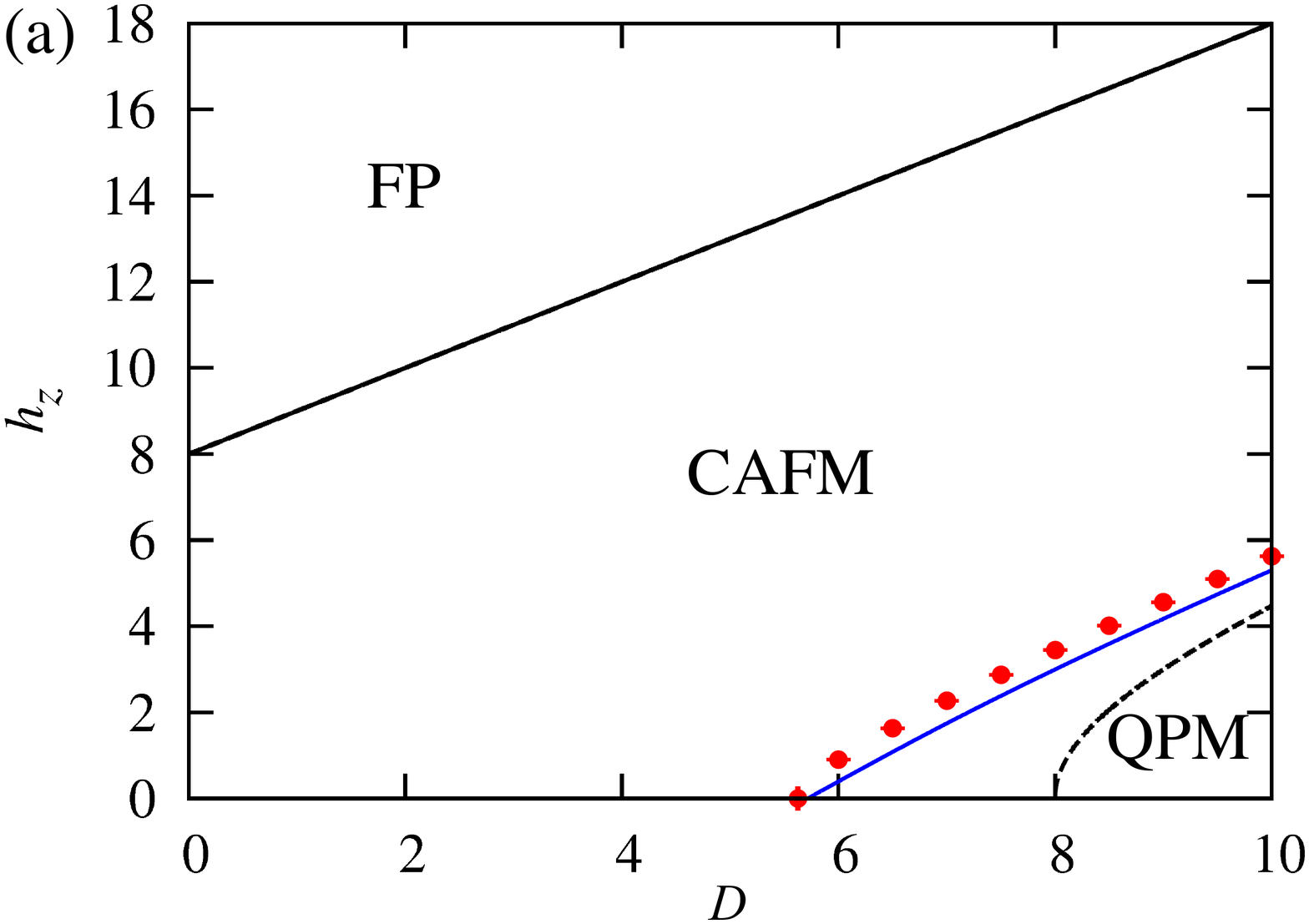} 
\includegraphics[width=8cm, trim=30 50 30 30,clip]{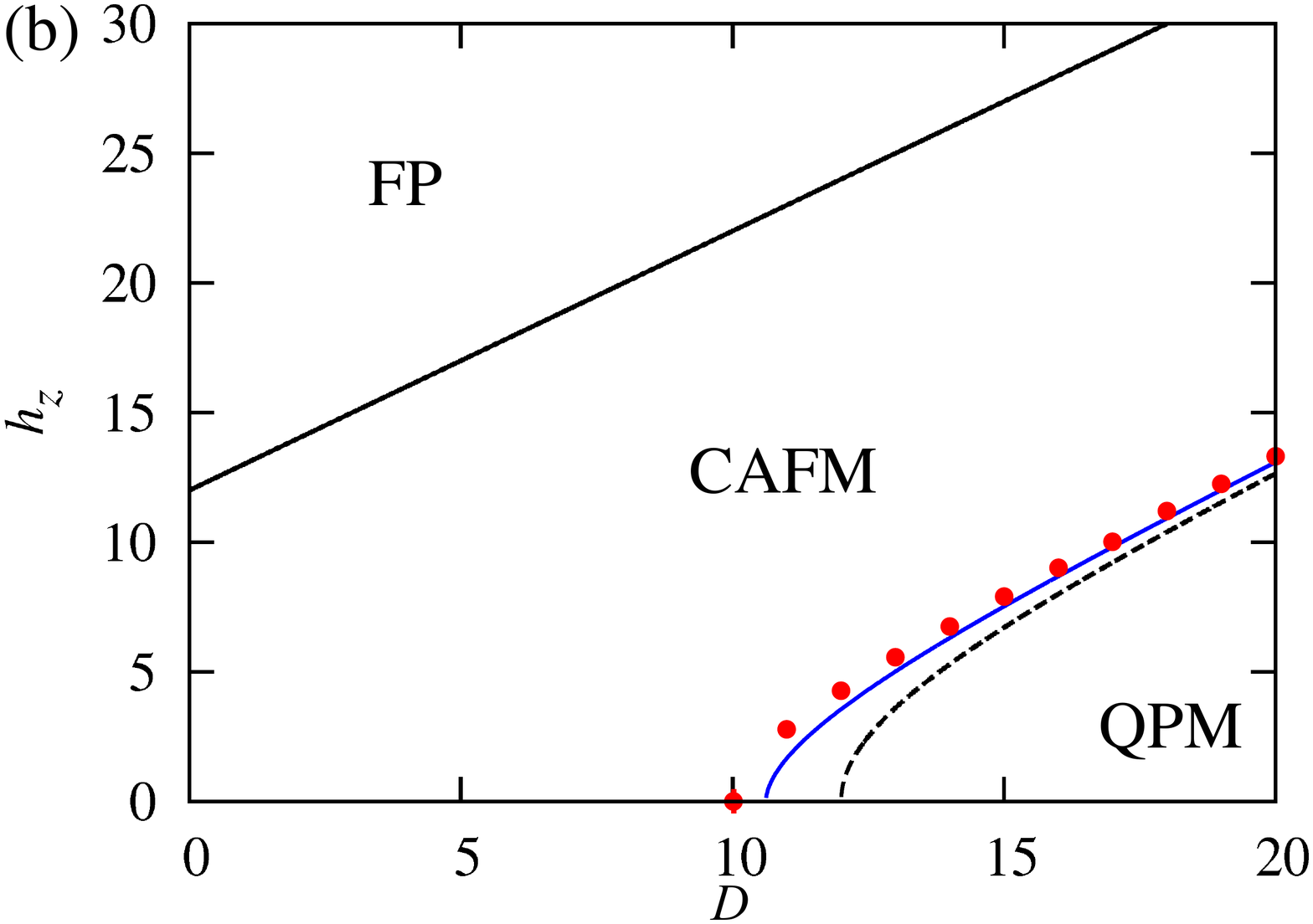} 
\caption{(Color online) Quantum phase diagram of ${\cal H}_H$ in (a) $d=2$ and (b) $d=3$. 
The solid line, dashed line and points between QPM and CAFM are the results obtained  from the LM, HP and QMC approaches, respectively.
For the  QMC approach we use the modified finite-size scaling that is described in the text as well as the gap that is obtained from the quasiparticle dispersion
to determine the QPM-CAFM phase boundary.
} 
\label{fig:PDs} 
\end{figure} 

\begin{figure}[hbt] 
\includegraphics[angle=0,width=9cm,trim=30 30 0 30,clip]{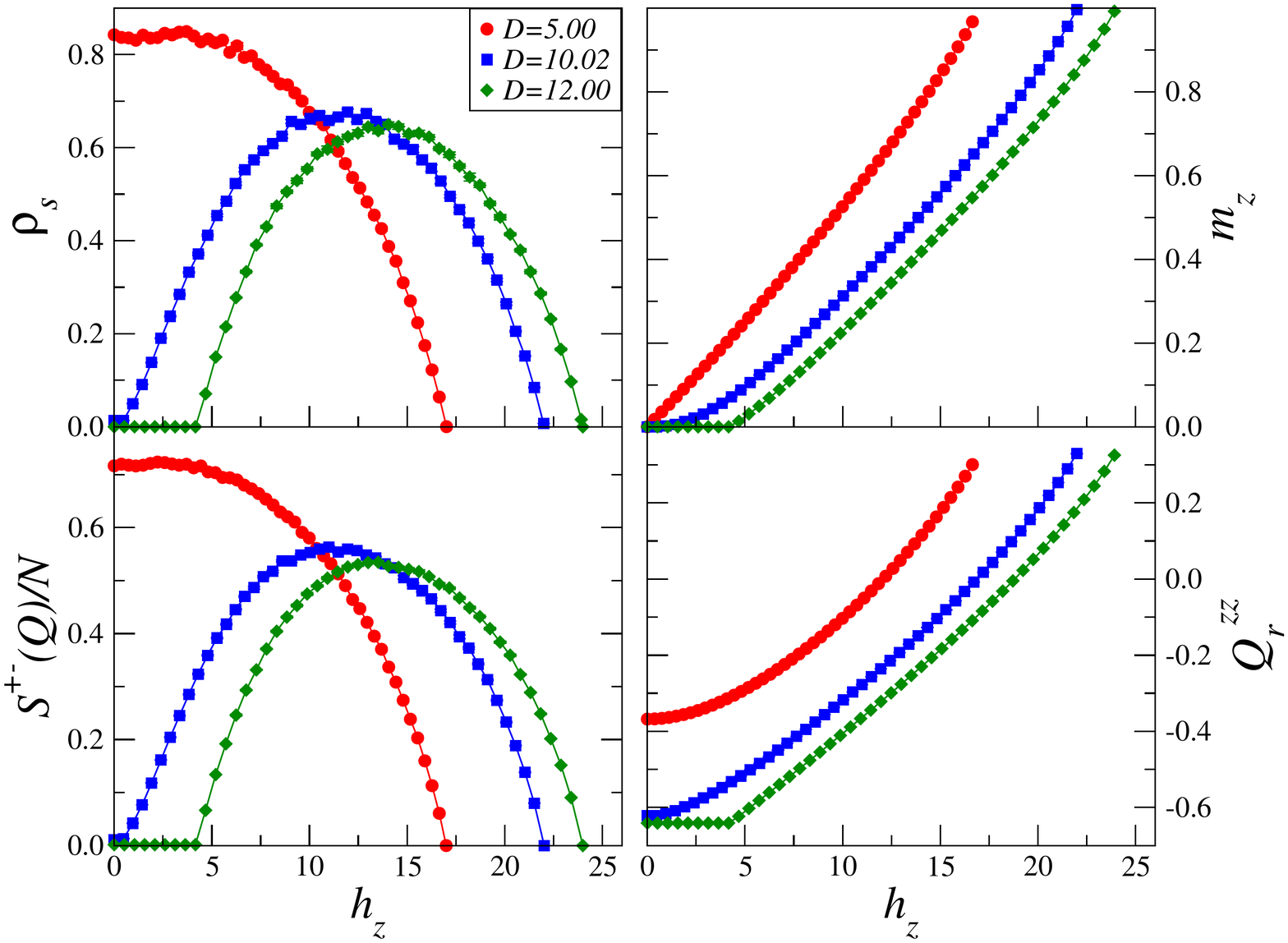} 
\caption{(Color online) The evolution of various characteristic observables with 
external magnetic field at three representative  values of $D$ as the ground 
state goes through the field driven quantum phase transitions discussed in 
the text. The data is for a finite cubic lattice of dimension $16\times 16\times 16$.}
\label{fig:groundstates} 
\end{figure} 

The quantum phase diagrams obtained with different methods: linear HP approximation, 
the LM approach and QMC simulations, are shown in Figs.~\ref{fig:PDs}. As it is expected from the comparisons between the quasiparticle dispersions 
obtained with the different methods in the QPM phase (see Fig.~\ref{fig:disp}),
the LM method produces a much better quantitative agreement with the QMC results than the linear HP approximation.
 
Fig.~\ref{fig:groundstates} shows the evolution of some observables that characterize the
ground state phases as the applied field is varied for three representative values 
of the single-ion anisotropy. For $D>D_c$, the ground state evolves from a QPM 
phase at low fields ($h_z < h_c$) to a CAFM phase at intermediate fields 
($h_c < h_z < h_s$) to a fully polarized phase at large fields.
  The uniform magnetization, $m_z$,  increase monotonically with the applied field. 
The {\em zz}-nematic order parameter, $Q^{zz}_{\bm r}$, also increases monotonically 
but from a negative to a positive value. Right above $h=h_c$, the magnetization 
$m_z$ increases with finite slope, but this slope vanishes at the O(2) QCP where $h_c(D_c)=0$. 
This result is consistent with the mean field theory described in the previous section which predicts that 
$m_z \propto (h_z - h_c(D))$ for finite $h_c(D)$ and small enough $h_z - h_c(D)$, while $m_z \propto h^3_z$
for $h_c=0$ and small enough $h_z$. These results are obtained by solving Eqs.\eqref{mfmin} near the 
O(2) QCP $(D=D_c, h_z=0)$. 

The stiffness and transverse structure
factor decrease monotonically with increasing $h_z$ for $D \ll D_c$. However, it is clear that the field dependence must be non-monotonic
for $D\geq D_c$,  because a finite critical field is required to induce the transition from the QPM to the ordered XY phase.
When the system is in the 
QPM phase, a critical field $h_c (D)$ is required to induce a finite amplitude of the XY order parameter, i.e., the mean field state of each spin becomes a linear 
combination of the states $| 0 \rangle_{\bm r}$ and $| 1 \rangle_{\bm r}$ for $h>h_c$. There is an optimal value of the magnetic field, $h_m (D)$, for which the weight of these two states
is roughly the same, leading to maxima of the order parameter (XY component of the local moment) and the spin stiffness, as it is shown in Fig.~\ref{fig:groundstates}. 
Finally, $\rho_s$ and $S^{+-}({\bm Q})$ vanish again at sufficiently strong applied field, $h_z \geq h_s(D)$, because the ground
state evolves to the fully polarized phase  with $m_z=1$, and ${\cal Q}^{zz}=1/3$. The exact boundary between
the CAFM and the FP phases is given by Eq.\eqref{hsat}. A simple continuity argument shows that the non-monotonic field dependence 
of $\rho_s$ and $S^{+-}({\bm Q})$ should persists for $D \lesssim D_c$ as it is clear from Fig.~\ref{fig:groundstates}. The ordering temperature should also
exhibit a similar non-monotonic field dependence, as we will see in the next section. This observation can be used to detect quantum magnets  that exhibit magnetic ordering at
$h_z=0$ , but are near the QCP, i.e., close to becoming quantum paramagnets.


\section{Finite-temperature results \label{ftr}}
\begin{figure}[htb] 
\includegraphics[angle=0,width=9cm,trim= 30 0 30 50 ,clip]{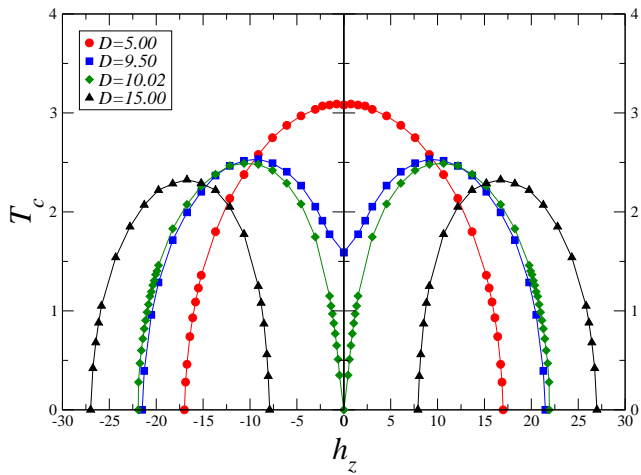}  
\caption{(Color online) The critical temperatures of the thermal phase transition into different ground states shown in Fig.~\ref{fig:PDs}(b).} 
\label{fig:mcdonalds} 
\end{figure} 
For three-dimensional systems, the CAFM phase survives up to a finite temperature $T_c(D,h_z)$ above which the system becomes a paramagnet 
via a second order classical phase transition that belongs to the O(2) universality class in dimension $d$. 
The second order transition is replaced by a Berezinskii-Kosterlitz-Thouless phase transition at $T=T_{BKT}$ when the system is two-dimensional. In this case, only quasi long range ordering survives at finite temperatures $T \leq T_{BKT}$.
Fig.~\ref{fig:mcdonalds} shows the field dependence of the  critical temperature, $T_c$, for some
representative values of $D$. 
$T_c$ is determined by exploiting the scale invariance of the stiffness at the critical point
with the finite-size scaling
 \begin{eqnarray}
    \rho_s(L, T )
    &\sim& L^{2-d} Y_{\rho_s}((T-T_c)L^{1/\nu}).
    \nonumber
\end{eqnarray}
The thermal transition out of the CAFM 
phase is driven by phase fluctuations of the order parameter and belongs to 
the $d=3$ O(2) universality class ($\nu \simeq 0.67$). At small
values of $D$, the system is dominated by the Heisenberg AFM interaction
and $T_c(h_z)$ decreases monotonically as a function of increasing $h_z$ to 
$T_c(h_{s})=0$ at the QMP-FP boundary. As $D$ increases, the spins acquire a significant $S^z=0$ (nematic) component and 
the resultant decrease in the local magnetization leads to a 
suppression of the critical temperature. As we explained in the previous section, the applied field increases the magnitude of the local moments
for $D \lesssim D_c$ and this effect leads to an accompanying increase in $T_c(h)$. At higher values of the
applied field, the spins acquire an increasing (ferromagnetic) component
along the field direction while the AFM-ordered component decreases beyond the optimal field $h_m (D)$. Consequently, the
 critical temperature  starts decreasing monotonically to $T_c(h_s)=0$ for $h > h_m (D)$. 
 For $D>D_c$, the system is in a QPM ground state at low
fields -- with the local spins being predominantly in the $S_z=0$ state --
 and $T_c=0$. A sufficiently strong external field induces a transition to
the  CAFM phase with $T_c \propto (h_z - h_c)^{2/3}$ for small enough $h_z - h_c$. The transition temperature increases initially 
as the magnitude of the local moments increase and eventually decreases 
as the moments acquire a dominant ferromagnetic component parallel to the 
applied field -- going to $T_c=0$ at $h_z=h_s$.

\section{Summary \label{sum}}

In summary, we have investigated the quantum phase diagram and the nature of the quantum phase transitions 
in the $S=1$ Heisenberg model with easy-plane single-ion anisotropy
and an external magnetic field. By using a generalized spin wave approach, we
showed that the low energy quasiparticle dispersion is qualitatively different at the phase boundary depending
on the presence or absence of an external field. This difference is reflected in the universality class of the 
underlying QCP and  has direct consequences on the low temperature 
behavior. The nature of the QPM-CAFM transition in the presence
and absence of an external field is directly confirmed by  using large scale QMC simulations and finite size scaling.

We have used two different analytical approaches to describe the QPM. By comparing the results of both approaches
against our QMC results, we have found important quantitative differences in the region near the O(2) QCP that signals the transition
to the CAFM phase. By ``quantitative differences" we are not referring to the already known critical behaviors predicted by both approaches, but to the phase
boundary $D_c (h_z)$ and the dispersion of the low-energy quasi-particle excitations. To make a clear distinction between these
two different aspects of the problem, we will discuss the critical behavior in the first place. It is clear that both analytical treatments
reproduce the correct critical behavior for $d=3$ up to logarithmic corrections, because  $d_c \geq 3$ for the QCPs  [O(2) and BEC] that appear in 
the quantum phase diagram of ${\cal H}_H$. The situation is different for $d=2$  because the upper critical dimension of the O(2) QCP is $d_c=3$.  
We note that the approach based on the inclusion of the  Lagrange multiplier and the saddle point approximation \eqref{speq}, becomes exact in the large ${\cal N} \to \infty $ limit (${\cal N}$ is the number of components of the order parameter of the broken symmetry state, i.e., ${\cal N}=2$ for the case under consideration) \cite{Sachdev99}. 
Since $\nu=1/(d-1)$ for ${\cal N} \to \infty$, the LM approach leads to a spin  gap that closes linearly  in $(D-D_c)$ for $d=2$  (see Fig.~\ref{fig:PDs}a). In contrast, the 
HP approach produces the  expected mean field exponent $\nu=1/2$.  Naturally, neither of these approaches can reproduce the correct value of the exponent $\nu$ [$\nu \simeq 0.67$ for the O(2) QCP in dimension ${\cal D}=2+1$] because $2 < d_c$. However, the LM approach can be systematically improved by including higher order corrections in $1/{\cal N}$.
The qualitative agreement for $d=3$ is not surprising because the effective dimension of the QCPs that appear in 
the quantum phase diagram of ${\cal H}_H$ is equal or higher than the upper critical dimension.

Since the limitations of the LM and HP approaches for describing the critical behavior of the O(2) QCP are already known, we have focused on the 
overall quantitative agreement for the phase boundary $D_c (h_z)$ and the dispersion of the low-energy quasi-particle excitations in comparison with the numerical results.
The very good  agreement between the LM and QMC results is rather surprising if we consider that it holds true even for $d=2$ (see Fig.~\ref{fig:disp} and \ref{fig:PDs}a).
Indeed, a similar treatment has been successfully applied to the quasi-one-dimensional organic quantum magnet known as  DTN \cite{Zapf06}. 
In this compound, the $S=1$ moments are provided by Ni$^{2+}$ ions which are arranged in a tetragonal lattice. The magnetic properties are well
described by the Hamiltonian \eqref{eq:H} with parameters $D=8.9 K$, $J_c=2.2 K$  
and $J_a=J_b=0.18 K$, where $J_\alpha$ denotes the strength of the Heisenberg 
exchange interaction along the different crystal axes. Once again, the introduction of a Lagrange multiplier to enforce the constraint \eqref{eq:constraint} leads to 
a critical field value of $\simeq 2$T, 
that is in very good agreement with the result of QMC simulations and with the experiments \cite{Zapf06,Zvyagin07}. 
In contrast, the linear HP approach incorrectly predicts that this compound should be magnetically ordered in absence of the applied magnetic field.
We note that the phase boundary obtained with the LM approach for $d=2$ (see Fig.~\ref{fig:PDs}a) remains quantitatively more accurate near the O(2) QCP even when  the next
(second) order corrections in 1/S are included  in the HP approach \cite{Hamer10}.
Our results then indicate that introducing a Lagrange multiplier for describing the low-energy physics of quantum paramagnets improves considerably the 
estimation of the spin gap and the quasiparticle dispersion. This improvement is particularly important for quantum paramagnets that have a small spin gap 
and consequently are close to the QCP that signals the onset of magnetic ordering. Since the Hamiltonian parameters are typically extracted from 
fits of the quasiparticle dispersion measured with INS, it is crucial to have a reliable approach for computing such dispersion. 
The  QMC method described in Sec.~\ref{quadisp} can only be applied to Hamiltonians that are free of the sign problem. However, the analytical approach described in Sec.~\ref{gsw}
is always applicable.

The numerical results were obtained in part using the computational resources of the National 
Energy Research Scientific Computing Center, which is supported by the Office of Science 
of the U.S. Department of Energy under Contract No. DE-AC02-05CH11231.


\begin{references}
\bibitem{note1}
We note that we are using the SB's as a convenient language for extending the large-$S$ approximation from SU(2) to SU(N) with $N=3$. Clearly, this approximation 
is completely different from the large-$N$ SB mean field theory introduced by Arovas and Auerbach \cite{Arovas88}.
\end{references}
\end{document}